\documentclass[lettersize,journal]{IEEEtran}
\usepackage{amsmath,amsfonts}
\usepackage{algorithmic}
\usepackage{algorithm}
\usepackage{array}
\usepackage[caption=false,font=normalsize,labelfont=sf,textfont=sf]{subfig}
\usepackage{textcomp}
\usepackage{stfloats}
\usepackage{url}
\usepackage{verbatim}
\usepackage{graphicx}
\usepackage{cite}

\begin{document}

\title{High Speed Emulation in a Vehicle-in-the-Loop Driving Simulator}

\author{Elliot Weiss and J. Christian Gerdes
\thanks{The authors are with the Department of Mechanical Engineering, Stanford University, Stanford, CA 94305 USA (e-mail: elliotdw@stanford.edu,
gerdes@stanford.edu).}}



\maketitle

\begin{abstract}
Rendering accurate multisensory feedback is critical to ensure natural user behavior in driving simulators. In this work, we present a virtual reality (VR)-based Vehicle-in-the-Loop (ViL) simulator that provides visual, vestibular, and haptic feedback to drivers in high speed driving conditions. Designing our simulator around a four-wheel steer-by-wire vehicle enables us to emulate the dynamics of a vehicle traveling significantly faster than the test vehicle and to transmit corresponding haptic steering feedback to the driver. By scaling the speed of the test vehicle through a combination of VR visuals, vehicle dynamics emulation, and steering wheel force feedback, we can safely and immersively run experiments up to highway speeds within a limited driving space. In double lane change and highway weaving experiments, our high speed emulation method tracks yaw motion within human perception limits and provides sensory feedback comparable to the same maneuvers driven manually.
\end{abstract}


\section{Introduction}
\IEEEPARstart{D}{riving} simulators are indispensable tools for understanding driver behavior and designing driver assistance systems. They enable the simulation of scenarios that are too dangerous to replicate in real life, such as overtaking on the highway. High speed scenarios are critical for testing driver assistance systems, as drivers must make quick and precise decisions to avoid other vehicles while keeping their vehicle stable. By designing a driving simulator specifically for high speed scenarios, we can better understand and assist drivers in risky situations on the highway.

A central challenge in the design of driving simulators is providing sensory feedback that elicits natural driving behavior from users. Beyond visual feedback, drivers rely on their internal perception of vehicle motion -- sensed by the vestibular system in the inner ear -- and the forces felt on their hands from the steering wheel for effective vehicle control. Repa \textit{et al.} observe the importance of lateral motion cues in studies of driver lane keeping in the presence of random wind gust disturbances, finding yaw motion to be particularly important in highly dynamic maneuvers \cite{repa1981influence}. Reymond \textit{et al.} additionally demonstrate that drivers rely on lateral acceleration cues to modulate speed while driving on curved roads \cite{reymond2001role}. Regarding haptic feedback, Toffin \textit{et al.} have studied driver adaptation to changes in haptic steering wheel feedback in driving simulators, finding that lateral vehicle control can be greatly influenced by changes in haptic feedback \cite{toffin2007role}. Cutlip \textit{et al.} explore the importance of steering feedback when switching from autonomous to manual control in driving simulators and report both fewer collisions and a greater awareness of control transitions among study participants when appropriate haptic feedback is present \cite{cutlip2021effects}. 

Moving platform driving simulators can provide multisensory feedback in high speed settings but typically at great cost and with limitations on the accelerations and angular velocities rendered to users. Yaw motion tends to be limited on hexapod-style platforms, which are commonly used in driving simulators. For example, Qazani \textit{et al.} run experiments with a hexapod simulator that can only rotate $\pm 20^{\circ}$ about the yaw axis \cite{qazani2019linear}. Moreover, great research effort has gone into the development of motion cueing algorithms that reproduce vehicle motion within a limited simulator workspace. Reymond and Kemeny present a common method of filtering reference accelerations into high and low frequency components for replication on the simulator via transient platform motion and tilting, respectively \cite{reymond2000motion}. This approach works well for smaller transient accelerations but fails to replicate motion once the platform can no longer accelerate in a given direction. In response to this limitation, simulator designers commonly build large facilities and use washout algorithms that slowly return the driver to the center of the simulator below their threshold of motion perception. Although more sophisticated motion cueing algorithms based on optimal and learning-based control have been developed for moving platforms, as shown by Garrett and Best \cite{garrett2010driving} and by Asadi \textit{et al.} \cite{asadi2019model}, the same limitations on rendering accelerations and rotations still exist. 

Vehicle-in-the-Loop (ViL) simulators represent an alternative approach to simulator design by placing the driver in a moving vehicle while viewing a virtually rendered driving scene. Bock \textit{et al.} developed an early ViL system using an augmented reality (AR) head-mounted display (HMD) to simulate other traffic participants \cite{bokc2007validation}. Similarly, Moussa \textit{et al.} describe an AR system using see-through video to run scenarios with other vehicles and variable weather conditions such as fog \cite{moussa2012augmented}. Several other ViL simulators, such as those presented by Karl \textit{et al.} \cite{karl2013driving}, Goedicke \textit{et al.} \cite{goedicke2018vr}, and Park \textit{et al.} \cite{park2020}, use virtual reality (VR) to visually immerse drivers. Contrasted with moving platform simulators, drivers experience the full motion of the vehicle across a wide range of frequencies in ViL and interact with the steering wheel of a real vehicle, enabling a more immersive experience. However, in parallel to motion platforms, ViL simulators have physical constraints on the range of motion available while testing based on the size of the track or proving ground in which tests take place. This limitation constrains the speed of the ViL test vehicle, as the space needed for safe testing increases with vehicle speed given the larger runway needed to accelerate to and decelerate from high speeds during an experiment.

We propose a method to overcome this limitation for ViL simulators that we call high speed emulation. In this method, we visually render the virtual vehicle seen by the driver through a VR HMD at a faster speed than the real vehicle. This idea is motivated by the observation that humans can perceive rotations and accelerations via their vestibular system but cannot easily distinguish between different velocities without a visual reference. To create a fully immersive experience while emulating high speeds, we use a four-wheel steer-by-wire (SBW) system to reproduce the yaw rate and lateral acceleration of the virtual vehicle driving at a higher speed while the real vehicle travels safely at a lower speed. Russell and Gerdes show that, with a four-wheel steer vehicle, the vehicle's motion can be altered to emulate a vehicle with different handling characteristics \cite{russell2015design}. Akar and Kalkkuhl use four-wheel steering to mimic the handling of buses and vans \cite{akar2008lateral}, and Russell and Gerdes demonstrate its use for reproducing the dynamics of driving on low friction surfaces \cite{russell2014low}. We apply similar techniques for emulating the lateral dynamics of a faster moving vehicle, thus providing appropriate vestibular feedback for the driver. Further, we use SBW technology to generate custom haptic feedback on the steering wheel regardless of the front road wheel angle, which may differ greatly from the driver's steering command when emulating high speeds. 

In this paper, we present a ViL driving simulator designed with a four-wheel SBW vehicle that provides realistic multisensory feedback to drivers. To the best of our knowledge, this is the first work showing a ViL system that uses four-wheel steer or SBW technology. We demonstrate that this driving simulator is capable of providing accurate visual, vestibular, and haptic feedback while emulating the motion of a high speed vehicle. Our high speed emulation approach enables experiments to take place within a limited testing area, greatly increasing safety while the driver is immersed in scenarios up to highway speeds. We further show that the four-wheel steering controller used to track lateral motion has stable error tracking dynamics. We verify our approach with ISO double lane change and highway weaving experiments at a range of lateral accelerations and speeds.

The remainder of this paper is structured in the following way. Section \ref{sys_architecture} provides a description of our ViL system architecture and an overview of high speed emulation. Section \ref{ref_dyns_model} presents the vehicle dynamics reference model used to simulate motion at a scaled speed. Section \ref{multisensory_FB} then describes our method for rendering visual, vestibular, and haptic feedback associated with the high speed reference model. Next, Section \ref{experiments} demonstrates our high speed emulation method with results from various experiments. Section \ref{conclusion} concludes with a discussion of limitations and future applications.

\section{System Architecture}
\label{sys_architecture}

\subsection{Hardware and Software}
The ViL system combines a virtual driving environment shown to the driver through a VR HMD with a full-sized four-wheel SBW vehicle. We use X1, a student-built research vehicle, as the physical testbed in this work. X1 is an all-electric steer-, throttle-, and brake-by-wire vehicle. The steering wheel and road wheels are mechanically decoupled, and a dedicated motor connected to the steering wheel provides haptic feedback to the driver. X1 has an OxTS RT4003 dual antenna GPS/INS unit on board that transmits high precision position, orientation, linear and rotational velocity, and acceleration measurements in real time. Table \ref{table_X1_params} lists X1's parameters.

\begin{table}
\begin{center}
    \caption{Relevant X1 Parameters}
    \label{table_X1_params}
    \begin{tabular}{ l  c  c  c }
    \hline
    Name & Symbol & Value & Units \\ 
    \hline
    mass & $m$ & 2000 & kg \\ yaw moment of inertia & $I_z$ & 2400 & kg$\cdot$m$^2$ \\ distance between CoM and front axle & $a$ & 1.52 & m \\ distance between CoM and rear axle & $b$ & 1.35 & m \\ track width & $d$ & 1.63 & m \\ steering ratio & SR & 15 & - \\ front tire stiffness & $C_f$ & 75000 & N/rad \\ rear tire stiffness & $C_r$ & 110000 & N/rad \\ friction coefficient & $\mu$ & 0.9 & - \\ maximum front steering angle & $\delta_{fmax}$ & 18 & deg \\ maximum rear steering angle & $\delta_{rmax}$ & 33 & deg \\
    \hline
    \end{tabular}
\end{center}
\end{table}

\begin{figure}[thpb]
    \centering
    \includegraphics[width=0.48\textwidth]{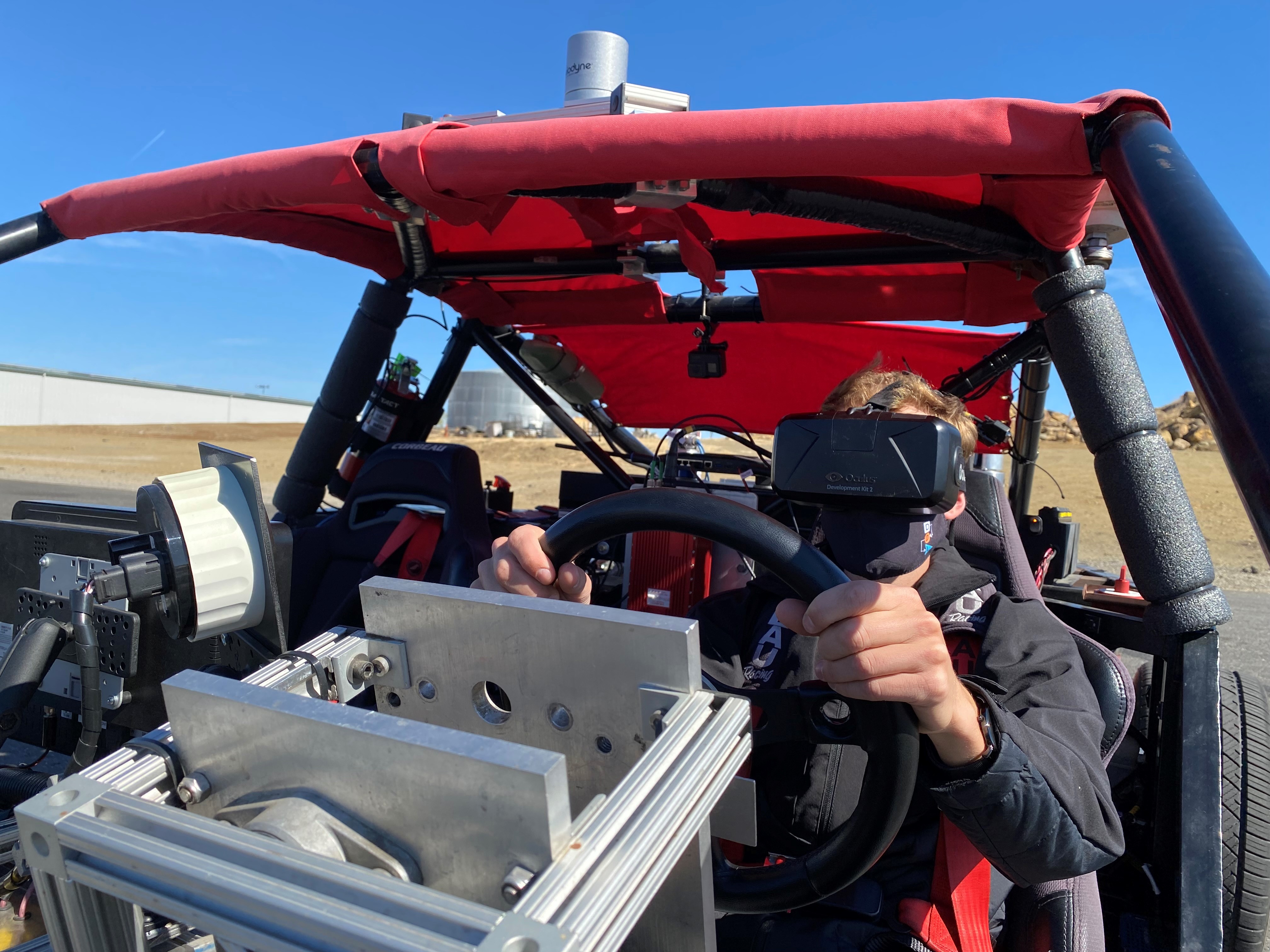}
    \caption{A driver in the ViL simulator on X1.}
    \label{figure_driver_in_vil}
\end{figure}

To create a stereo rendering of the virtual world for the driver, we use Virtual Test Drive (VTD), a proprietary driving simulation software package \cite{von2009virtual}. In VTD, we can set up safe experiments with other virtual road users in a variety of environments and update the simulation in real time based on measurements and computations on board X1. The virtual scene is displayed to the driver through an Oculus Rift Development Kit 2 VR HMD. The virtual camera's roll, pitch, and yaw angles update according to the orientation of the driver's head, which is computed as a fusion of three-axis accelerometer, gyroscope, and magnetometer measurements on the HMD, following work by LaValle \textit{et al.} \cite{lavalle2014head}. The Robot Operating System (ROS) software package facilitates real time communication between the VTD simulation, the Oculus Rift HMD, and X1's sensors and actuators through multiple on board computers. Fig. \ref{figure_driver_in_vil} shows a front view of our ViL platform.

\subsection{Control System Overview}
\begin{figure}[thpb]
    \centering
    \includegraphics[width=\linewidth, trim={0.5cm 1.5cm 1.5cm 0.5cm}, clip]{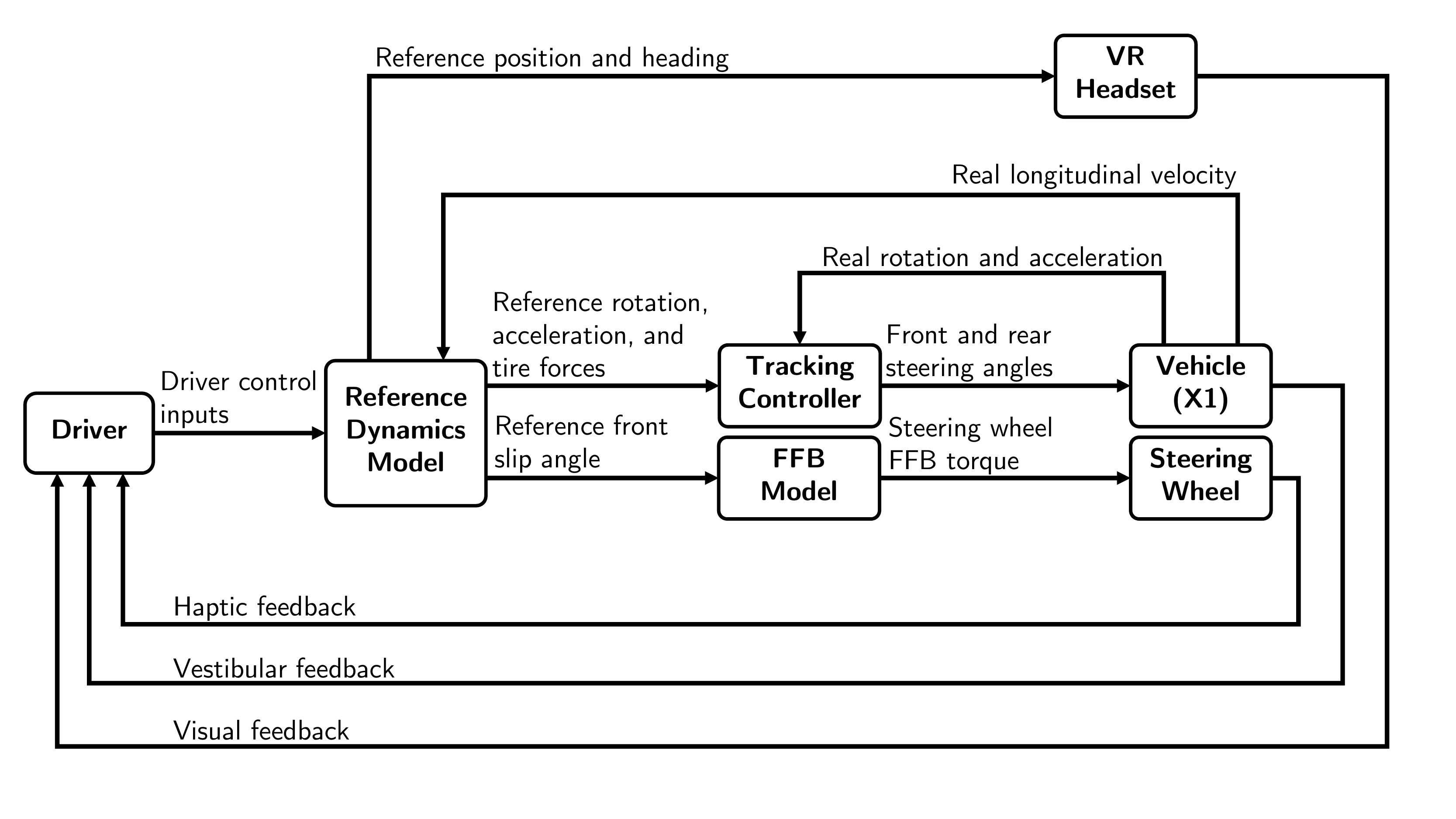}
    \caption{Overview of the system used to emulate high speeds.}
    \label{figure_control_system_overview}
\end{figure}

The flow of signals for high speed emulation on X1 is depicted as a block diagram in Fig. \ref{figure_control_system_overview}. Beginning with the driver, sensors on X1 measure their steering, throttle, and brake commands for input to a reference dynamics model. The reference model additionally takes as input a measurement of X1's current speed. This speed is scaled by a constant factor in the reference model, enabling the model to simulate the motion of X1 as if it were traveling at a higher speed. The reference model simulates the tire forces and motion of a vehicle traveling at the scaled  speed. State information from the reference model feeds into three different modules that provide sensory feedback to the driver. The position and heading from the reference model update the motion of the vehicle in the virtual world, providing the driver visual feedback associated with the higher speed vehicle motion. The yaw rate and lateral acceleration of the reference model are tracked using four-wheel steering on X1 to render accurate vestibular feedback to the driver. The tracking controller uses tire forces generated in the reference model as feedforward terms and the real rotation and acceleration of X1 for feedback. Finally, the front slip angle from the reference model is an input to a force feedback model used to compute the steering wheel torque needed for appropriate haptic feedback. The details of the reference model and the methods used to provide multisensory feedback to the driver are described in more detail in the following sections.

\section{Reference Dynamics Model}
\label{ref_dyns_model}

\begin{figure}[thpb]
    \centering
    \includegraphics[width=0.48\textwidth]{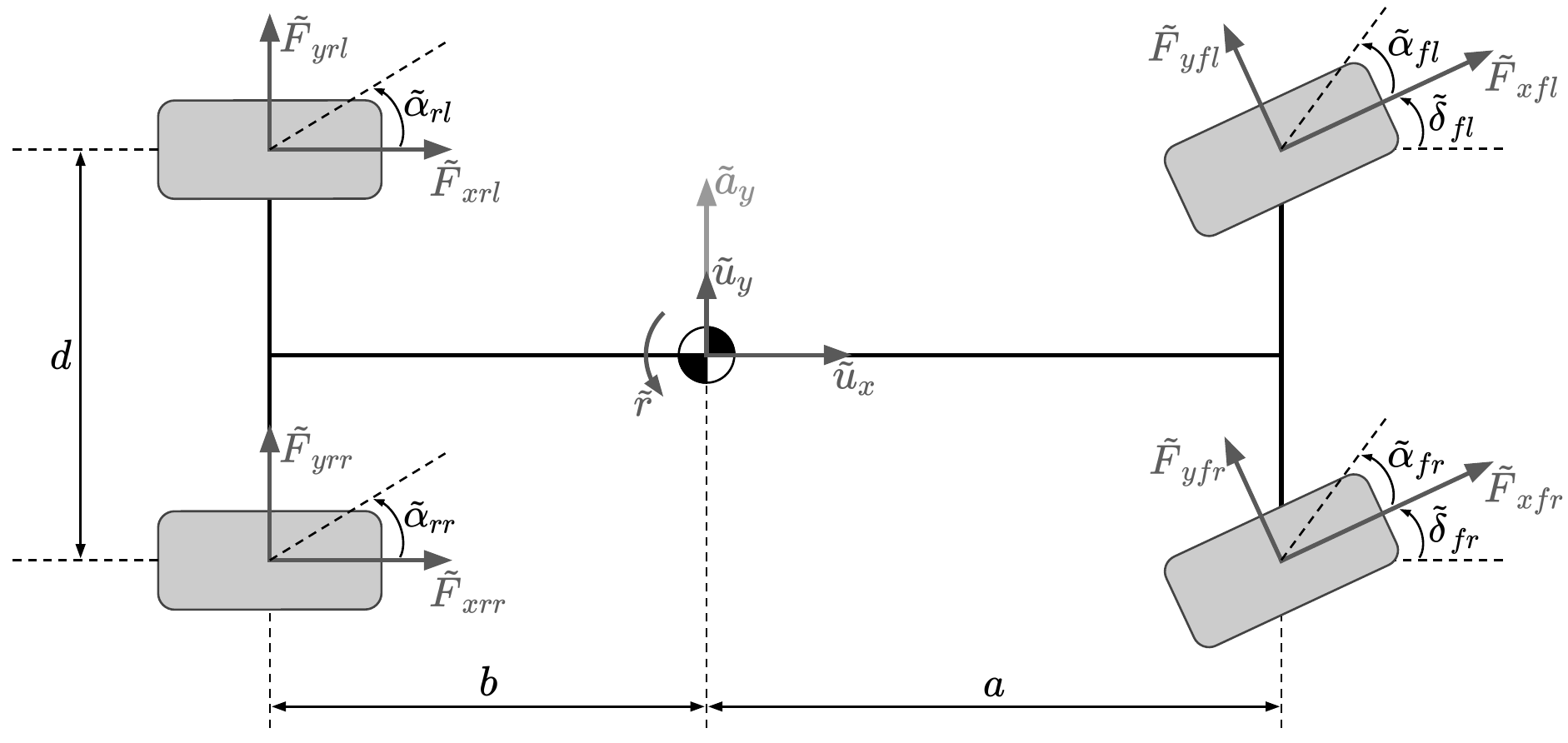}
    \caption{A schematic diagram of the double track vehicle model used to simulate the high speed reference dynamics.}
    \label{figure_double_track_model}
\end{figure}

The reference dynamics model is central to this work, as the vehicle state computed by this model directly contributes to the rendering of visual, vestibular, and haptic feedback. We base our double track reference model on work by Russell and Gerdes, who use a similar model for modifying the handling dynamics of a four-wheel steer vehicle \cite{russell2015design}. For clarity, we print the variables computed within the reference model with a tilde symbol over top. Fig. \ref{figure_double_track_model}. shows the reference model diagrammatically.

The speed in the high speed reference model is the longitudinal velocity measured on X1, $u_x$, scaled by a constant
\begin{equation}
    \Tilde{u}_x = fu_x,
    \label{ref_model_ux}
\end{equation}
where $f > 1$ is a user defined scaling factor. Directly scaling the speed enables the driver to maintain longitudinal control of the vehicle starting from rest.

\subsection{Lateral Vehicle Dynamics}
The lateral motion of the vehicle is more sensitive to steering inputs at higher speeds. This comes from the observation that it only requires a small steering input to change lanes on the highway, while large steering angles are needed to move the vehicle the same amount laterally at low speeds, such as in a parking lot. To simulate the lateral dynamics, we use two states: yaw rate $\Tilde{r}$ and lateral velocity $\Tilde{u}_y$. The equations of motion for these states are written in terms of the total moment $\Tilde{M}_z$ and total lateral force $\Tilde{F}_y$ as
\begin{equation}
\begin{aligned}
    \dot{\Tilde{r}} &= \frac{\Tilde{M}_z}{I_{z}} \\ 
    \dot{\Tilde{u}}_y &= \frac{\Tilde{F}_y}{m}-\Tilde{r}\Tilde{u}_x. \label{ref_model_EOMs}
\end{aligned}
\end{equation}
We compute the total moment and lateral force on the vehicle in terms of forces at each tire's contact patch with the ground
\begin{equation}
\begin{aligned}
    \Tilde{M}_z &= a(\Tilde{F}_{xfl}\sin{\Tilde{\delta}_{fl}}+\Tilde{F}_{yfl}\cos{\Tilde{\delta}_{fl}})~... \\           &+a(\Tilde{F}_{xfr}\sin{\Tilde{\delta}_{fr}}+\Tilde{F}_{yfr}\cos{\Tilde{\delta}_{fr}})-b(\Tilde{F}_{yrl}+\Tilde{F}_{yrr})~... \\
    &-\frac{d}{2}(\Tilde{F}_{xfl}\cos{\Tilde{\delta}_{fl}}-\Tilde{F}_{yfl}\sin{\Tilde{\delta}_{fl}})~... \\ &+\frac{d}{2}(\Tilde{F}_{xfr}\cos{\Tilde{\delta}_{fr}}-\Tilde{F}_{yfr}\sin{\Tilde{\delta}_{fr}}-\Tilde{F}_{xrl}+\Tilde{F}_{xrr}) \\
    \Tilde{F}_y &= \Tilde{F}_{xfl}\sin{\Tilde{\delta}_{fl}}+\Tilde{F}_{yfl}\cos{\Tilde{\delta}_{fl}}~... \\           &+\Tilde{F}_{xfr}\sin{\Tilde{\delta}_{fr}}+\Tilde{F}_{yfr}\cos{\Tilde{\delta}_{fr}}+\Tilde{F}_{yrl}+\Tilde{F}_{yrr}.
    \label{ref_model_MzandFy}
\end{aligned}
\end{equation}
Summing moments and forces in this way is useful, as we use $\Tilde{M}_z$ and $\Tilde{F}_y$ to compute feedforward forces in the tracking controller. The front road wheel angles directly relay the driver's steering command as $\Tilde{\delta}_{fr} = \Tilde{\delta}_{fl} = \delta_{hw}/\text{SR}$, where $\delta_{hw}$ is the measured hand wheel angle. 

Derived from 2D rigid body kinematics, the lateral acceleration at the vehicle's center of mass is
\begin{equation}
    \Tilde{a}_y = \Dot{\Tilde{u}}_y+\Tilde{r}\Tilde{u}_x.
    \label{ref_model_ay}
\end{equation}
These two terms arise from modeling the vehicle as a translating and rotating rigid body that experiences both linear and centripetal acceleration components.

\subsection{Tire Model}
To determine the forces used in the equations of motion, we need to compute slips and forces for each tire based on vehicle state and control input values. We use a coupled slip tire model to simultaneously calculate the longitudinal and lateral forces on each tire. The longitudinal slip $\Tilde{\sigma}_x$ and lateral slip $\Tilde{\sigma}_y$ have a combined magnitude of $\Tilde{\sigma} = \sqrt{\Tilde{\sigma}_x^2+\Tilde{\sigma}_y^2}$. The slips are inputs to a coupled slip form of Pacejka's brush tire model \cite{pacejka2012tire}, which computes the forces on each tire as
\begin{equation}
    \Tilde{F}_x = \frac{\Tilde{\sigma}_x}{\Tilde{\sigma}}\Tilde{F},~ \Tilde{F}_y = \frac{\Tilde{\sigma}_y}{\Tilde{\sigma}}\Tilde{F},
    \label{ref_model_tireFyFx_from_F}
\end{equation}
where
\begin{equation}
    \Tilde{F} = 
    \begin{cases}
    \Tilde{\sigma} C -\frac{\Tilde{\sigma}^2C^2}{3\mu F_z}+\frac{\Tilde{\sigma}^3C^3}{27\mu^2F_z^2},& \Tilde{\sigma} \leq \sigma_{sl} \\
    \mu F_z,& \Tilde{\sigma} > \sigma_{sl},
    \end{cases}
    \label{ref_model_coupled_slip_fiala}
\end{equation}
and $\sigma_{sl} = (3\mu F_z/C)$ is the peak slip magnitude before the tire is fully sliding. We assume an isotropic cornering stiffness $C$ on each tire, defined for the front and rear axles in Table \ref{table_X1_params}. 

Slips arise from the deformation of rubber along each tire's $x$ and $y$ axes. We compute longitudinal slip $\Tilde{\sigma}_x$ from the driver's longitudinal force input on each wheel by inverting the coupled slip model in (\ref{ref_model_tireFyFx_from_F}) and (\ref{ref_model_coupled_slip_fiala}) with $\Tilde{\sigma}_y = 0$, following the approach presented by Russell and Gerdes in \cite{russell2015design}. The driver's throttle and braking commands linearly map to longitudinal forces on each tire using a symmetric force distribution between the left and right tires of each axle. Lateral tire slip is calculated as a function of longitudinal tire slip following the geometric relationship $\Tilde{\sigma}_y = (\Tilde{\sigma}_x-1)\tan{\Tilde{\alpha}}$, where $\Tilde{\alpha}$ is the slip angle. We compute $\Tilde{\alpha}$ for each tire as the angle between its velocity vector and its centerline, for example on the front left tire
\begin{equation}
\begin{aligned}
    \Tilde{\alpha}_{fl} &= \tan ^{-1} \left(\frac{\Tilde{u}_y+a\Tilde{r}}{\Tilde{u}_x-\Tilde{r}\frac{d}{2}}\right)-\Tilde{\delta}_{fl}.
    \label{ref_model_slip_angles}
\end{aligned}
\end{equation}
This method provides sufficient detail for computing tire slips and forces in our double track reference model. Other methods that capture tire nonlinearities, for example those that incorporate wheel speed dynamics, could also be used to model tire forces in this approach.

\section{Multisensory Feedback}
\label{multisensory_FB}

We use the states and forces computed in the reference model to render cohesive visual, vestibular, and haptic feedback to the driver. Our goal is to fully immerse the driver in the high speed setting while driving at a lower speed.

\subsection{Visual Feedback}

Perhaps more than any other sense, humans use visual feedback to stay on the road and avoid collisions with other road users. When emulating high speeds, it is critical to give the driver the visual perception of moving at the faster reference speed $\Tilde{u}_x$. The driver should also see lateral vehicle motion according to the reference yaw rate $\Tilde{r}$ and lateral velocity $\Tilde{u}_y$.

\begin{figure}[thpb]
    \centering
    \includegraphics[width=0.48\textwidth]{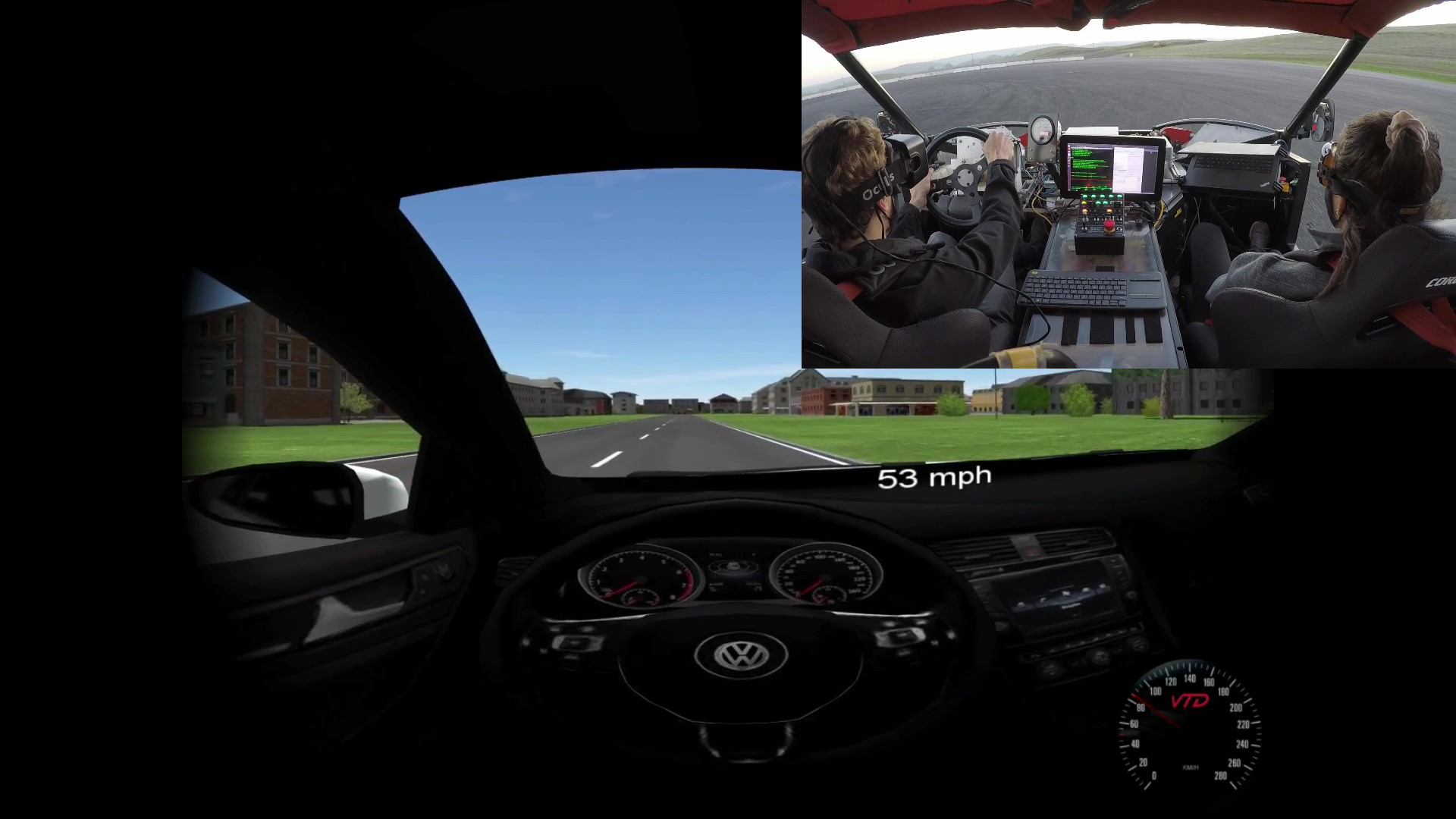}
    \caption{The real and virtual worlds overlaid during a test.}
    \label{figure_real_virtual_overlay}
\end{figure}

The heading and position coordinates of the virtual vehicle evolve according to
\begin{equation}
\begin{aligned}
    \dot{\Tilde{\psi}} &= \Tilde{r} \\ 
    \dot{\Tilde{E}} &= -\Tilde{u}_x \sin{\Tilde{\psi}} -\Tilde{u}_y \cos{\Tilde{\psi}} \\ 
    \dot{\Tilde{N}} &= ~~\Tilde{u}_x \cos{\Tilde{\psi}} -\Tilde{u}_y \sin{\Tilde{\psi}},
 \label{ref_model_pos_rot}
\end{aligned}
\end{equation}
where $\Tilde{\psi}$ is the reference vehicle's heading angle relative to north, and $(\Tilde{E}, ~\Tilde{N})$ are the reference vehicle's position coordinates in an inertial reference frame. No matter the motion of the real vehicle, the rotations and velocities of the virtual vehicle seen by the driver in the VR HMD match the dynamics of the reference model. In addition, the reference speed $\Tilde{u}_x$ is shown on a heads up display in the virtual vehicle. Fig. \ref{figure_real_virtual_overlay} shows the driver's virtual view during a high speed maneuver overlaid with a first-person view inside X1. 

\subsection{Vestibular Feedback}

\begin{figure}[thpb]
    \centering
    \includegraphics[width=0.48\textwidth]{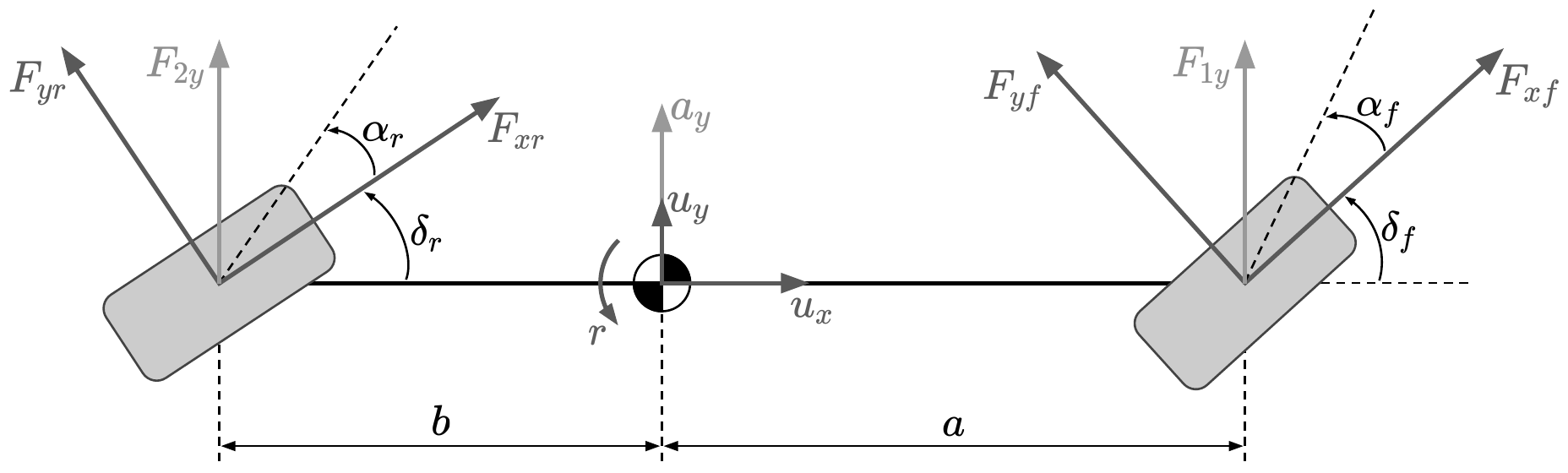}
    \caption{The single track model used for steering control and haptic feedback.}
    \label{figure_single_track_model}
\end{figure}

Drivers take cues from their vehicle's motion to maneuver safely on the road. We track yaw rate $\Tilde{r}$ and lateral acceleration $\Tilde{a}_y$, so that the driver receives vestibular feedback consistent with the high speed motion of the reference vehicle. We have developed a controller that computes the front and rear steering inputs needed to track these values. Note that the longitudinal inputs to the vehicle are controlled directly by the driver through throttle and brake commands. We base the steering controller on a planar four-wheel steer, single track vehicle model that captures the relevant dynamics on X1, shown in Fig. \ref{figure_single_track_model}. Note that the haptic feedback method in Section \ref{section_haptic_feedback} also uses the single track vehicle model.

\subsubsection{Control Law}
We implement a feedforward-feedback controller to compute front and rear steering commands. Together, these two actuators track the two reference states of interest. Similar to the method proposed by Russell and Gerdes for handling emulation on four-wheel steer vehicles, the controller first computes body-fixed tire forces $F_{1y}$ and $F_{2y}$ and then converts these forces into steering commands on each axle \cite{russell2015design}. In terms of these tire forces, the equations of motion for the real vehicle are
\begin{equation}
\begin{aligned}
    \dot{r} &= \frac{aF_{1y}-bF_{2y}}{I_{z}} \\ 
    \dot{u}_y &= \frac{F_{1y}+F_{2y}}{m}-ru_x. 
    \label{controller_EOMs}
\end{aligned}
\end{equation}

The lateral acceleration of the real vehicle $a_y$ has the same linear and rotational components as the reference vehicle, shown in (\ref{ref_model_ay}). To track the reference lateral acceleration, we set $\Tilde{a}_y = a_y$ and solve for $\dot{u}_{ydes}$, the desired derivative of lateral velocity on the vehicle, as
\begin{equation}
    \dot{\Tilde{u}}_y + \Tilde{r}\Tilde{u}_x = \dot{u}_y + ru_x \Rightarrow \dot{u}_{ydes} = \dot{\Tilde{u}}_y + \Tilde{r}\Tilde{u}_x - ru_x.
    \label{controller_set_ay_equal}
\end{equation}
Since we track the reference yaw rate at a slower speed on the real vehicle, $|\Tilde{r}\Tilde{u}_x| > |ru_x|$ during lateral maneuvers. To compensate for this decrease in the rotational component of lateral acceleration, the linear component increases on the vehicle. Thus, the desired $\Dot{u}_{ydes}$ tends to exceed the reference $\Dot{\Tilde{u}}_y$ in magnitude. This increase in $\dot{u}_y$ leads to exaggerated lateral motion, which can be achieved with the four-wheel steer capability of X1. For use in the tracking controller, we integrate the value of $\dot{u}_{ydes}$ in (\ref{controller_set_ay_equal}) to compute the desired lateral velocity
\begin{equation}
    u_{ydes} = \int \left(\dot{\Tilde{u}}_y + \Tilde{r}\Tilde{u}_x - ru_x\right)\, dt.
    \label{controller_integrate_uydesdot}
\end{equation}
Critically, this integration introduces a pure integrator into the closed-loop system, and the system is not stable when only using proportional feedback. Due to the pure integrator, steady-state errors in one variable become ramp errors in others. On a four-wheel SBW vehicle, a small misalignment in the steering system may introduce a steady-state yaw rate error, which causes the lateral velocity error to grow and the vehicle to drift sideways during tests.

To resolve this issue, we use the following control law with both proportional and integral feedback terms:
\begin{equation}
\begin{aligned}
    F_{1y} = \frac{b\Tilde{F}_y}{L}+\frac{\Tilde{M}_z}{L}&+K_{1r}e_r+K_{1rI}\int e_r\,dt~... \\
    &+K_{1uy}e_{uy}+K_{1uyI}\int e_{uy}\,dt \\ 
    F_{2y} = \frac{a\Tilde{F}_y}{L}-\frac{\Tilde{M}_z}{L}&+K_{2r}e_r+K_{2rI}\int e_r\,dt~... \\
    &+K_{2uy}e_{uy}+K_{2uyI}\int e_{uy}\,dt, 
    \label{controller_F12y_ctlLaw}
\end{aligned}
\end{equation}
where the error states used in feedback are defined as
\begin{equation}
\begin{aligned}
    e_r &= \Tilde{r}-r \\ 
    e_{uy} &= u_{ydes}-u_y. 
    \label{controller_define_er_euy}
\end{aligned}
\end{equation}
The feedforward terms use $\Tilde{F}_y$ and $\Tilde{M}_z$ to match the desired lateral forces on the front and rear axles. The feedback terms correct for unmodeled dynamics and ensure system stability by accounting for steady-state errors with integral feedback. 

\subsubsection{Stability Analysis}
Although the desired lateral forces computed in (\ref{controller_F12y_ctlLaw}) are not perfectly achieved given model mismatch in the conversion from forces to steering angles, we can analyze the closed-loop system to understand the stability of the error dynamics. If the system is stable, errors in tracking reference states will decay over time. Taking the derivative of error states in (\ref{controller_define_er_euy}) and substituting the dynamics in (\ref{ref_model_EOMs}) and  (\ref{controller_EOMs}) and the definition of $u_{ydes}$ in (\ref{controller_integrate_uydesdot}), we get the following expressions for the dynamics of the yaw rate and lateral velocity error states in terms of moments and forces:
\begin{equation}
\begin{aligned}
    \dot{e}_r &= \frac{\Tilde{M}_z-aF_{1y}+bF_{2y}}{I_z} \\ 
    \dot{e}_{uy} &= \frac{\Tilde{F}_y-F_{1y}-F_{2y}}{m}.
    \label{controller_define_erdot_euydot}
\end{aligned}
\end{equation}
We also analyze the dynamics of lateral acceleration error $e_{ay} = \Tilde{a}_y-a_y$. Using the definition of lateral acceleration in (\ref{ref_model_ay}) and the equations of motion in (\ref{ref_model_EOMs}) and (\ref{controller_EOMs}), the lateral acceleration error dynamics are
\begin{equation}
    \dot{e}_{ay} = \frac{\dot{\Tilde{F}}_y-\dot{F}_{1y}-\dot{F}_{2y}}{m} = \Ddot{e}_{uy}. 
    \label{controller_define_eaydot}
\end{equation}
By calculating the desired lateral velocity required to track the reference lateral acceleration, the dynamics on $e_{ay}$ are simply the derivative of the dynamics on $e_{uy}$. We only track the desired lateral velocity $u_{ydes}$ to match the reference lateral acceleration, and thus lateral velocity error dynamics operate independently of yaw rate error dynamics. To form a closed-loop system in terms of error states, we substitute the control law from (\ref{controller_F12y_ctlLaw}) into the derivatives of error dynamics in (\ref{controller_define_erdot_euydot}) and (\ref{controller_define_eaydot}). Additionally differentiating the expression for $\dot{e}_r$ in (\ref{controller_define_erdot_euydot}), we derive a linear 4th order system of the form
\begin{equation}
    \begin{bmatrix} \Ddot{e}_r \\ \dot{e}_r \\ \dot{e}_{ay} \\ \dot{e}_{uy} \end{bmatrix} = \begin{bmatrix} K_1 & K_2 & K_3 & K_4 \\ 1 & 0 & 0 & 0 \\ K_5 & K_6 & K_7 & K_8 \\ 0 & 0 & 1 & 0 \end{bmatrix} \begin{bmatrix} \dot{e}_r \\ e_r \\ e_{ay} \\ e_{uy} \end{bmatrix},
    \label{controller_error_dynamics_matrix}
\end{equation}
where matrix elements $K_1$ through $K_8$ are defined in terms of the proportional and integral feedback gains in the control law. We can select these gains such that all eigenvalues of the dynamics matrix have negative real parts, ensuring an asymptotically stable system. Note that, by including $\dot{e}_r$, we achieve stable error dynamics on yaw acceleration $\dot{r}$, in addition to yaw rate, lateral acceleration, and lateral velocity. Within the set of gains resulting in system stability, we choose values that result in good transient tracking behavior on the vehicle. The definitions of matrix elements and gain values tuned in our experiments are shown in Table \ref{table_gains}.

\begin{table}
\begin{center}
    \caption{Feedback Gain Values}
    \label{table_gains}
    \renewcommand{\arraystretch}{1.5} 
    \begin{tabular}{ c  c  c  c }
    \hline
    Matrix Element & Definition & Value & Units \\ 
    \hline
    $K_1$ & $\frac{-aK_{1r}+bK_{2r}}{I_z}$ & -24.9 & $\frac{1}{s}$ \\ $K_2$ & $\frac{-aK_{1rI}+bK_{2rI}}{I_z}$ & -74.7 & $\frac{1}{s^2}$ \\ $K_3$ & $\frac{-aK_{1uy}+bK_{2uy}}{I_z}$ & 1.2 & $\frac{1}{m \cdot s}$ \\ $K_4$ & $\frac{-aK_{1uyI}+bK_{2uyI}}{I_z}$ & 3.6 & $\frac{1}{m \cdot s^2}$ \\ $K_5$ & $\frac{-K_{1r}-K_{2r}}{m}$ & 3.0 & $\frac{m}{s}$ \\ $K_6$ & $\frac{-K_{1rI}-K_{2rI}}{m}$ & 9.0 & $\frac{m}{s^2}$ \\ $K_7$ & $\frac{-K_{1uy}-K_{2uy}}{m}$ & -15.0 & $\frac{1}{s}$ \\ $K_8$ & $\frac{-K_{1uyI}-K_{2uyI}}{m}$ & -45.0 & $\frac{1}{s^2}$ \\
    \hline
    \end{tabular}
\end{center}
\end{table}

\subsubsection{Computing Steering Commands}
We convert the lateral forces computed in (\ref{controller_F12y_ctlLaw}) to road wheel angle inputs on each axle $\delta_f$ and $\delta_r$ for lateral control on X1. To make this conversion, we first transform the body-fixed forces to tire-fixed coordinates. Using the steering angles shown in Fig. \ref{figure_single_track_model}, the tire-fixed forces are
\begin{equation}
\begin{aligned}
    F_{yf} &= \frac{F_{1y}-F_{xf}\sin{\delta_f}}{\cos{\delta_f}} \\
    F_{yr} &= \frac{F_{2y}-F_{xr}\sin{\delta_r}}{\cos{\delta_r}},
    \label{controller_body_to_tire_forces}
\end{aligned}
\end{equation}
where $F_{xf}$ and $F_{xr}$ are the driver's longitudinal control inputs to the vehicle's drive motor and brake-by-wire system.

To determine the steering angles that will produce the lateral tire forces $F_{yf}$ and $F_{yr}$, we invert the brush tire model in (\ref{ref_model_tireFyFx_from_F}) and (\ref{ref_model_coupled_slip_fiala}) with $\sigma_x = 0$ to compute lateral slip angles $\alpha_f$ and $\alpha_r$, assuming minimal longitudinal slip and tire forces lumped together in the single track model. The accelerations achieved while emulating high speeds coincide with forces in the nonlinear regime of the tires, thus requiring an inversion of the brush tire model. We lastly use a single track definition of slip angles to calculate steering angles using the computed $\alpha_f$ and $\alpha_r$ values and measurements of velocity states on X1
\begin{equation}
\begin{aligned}
    \delta_f &= -\alpha_f + \tan ^{-1} \left(\frac{u_y+a r}{u_x}\right) \\ 
    \delta_r &= -\alpha_r + \tan ^{-1} \left(\frac{u_y-b r}{u_x}\right).
    \label{controller_steering_from_slip}
\end{aligned}
\end{equation}
With these two steering inputs, the controller tracks the reference yaw rate $\Tilde{r}$ and lateral acceleration $\Tilde{a}_y$ while accounting for nonlinearities inherent to the road wheel angles and tire forces of the vehicle.

\subsubsection{Actuator Saturation}
When emulating high speeds, the rear wheels tend to steer in the same direction as the front wheels. This enables the significant lateral motion needed to track the reference lateral acceleration on the vehicle. To track the reference yaw rate, the front wheels usually steer with slightly greater magnitude than the rear wheels during transient maneuvers. As a result, the front axle, which has about half the maximum steering angle of the rear axle, reaches its limit first.

To account for front steering saturation, we follow a similar approach to that outlined by Russell and Gerdes in the case of rear axle steering saturation during low friction emulation \cite{russell2015design}. When $\delta_f$ exceeds its limit $\delta_{fmax}$, we can only modify rear steering for control and choose to prioritize tracking yaw rate at the expense of lateral acceleration. If we kept tracking lateral acceleration, the rear wheels would over-rotate the vehicle, causing it to change heading significantly and risk driving out of a safe testing space. In contrast, tracking yaw rate maintains a real world heading angle that closely matches that of the virtual vehicle.

To track reference yaw rate when the front steering angle saturates, we first estimate the current front lateral force by computing the front slip angle using measured states on the vehicle, solving for $\alpha_f$ in (\ref{controller_steering_from_slip}). We then compute the corresponding lateral force with the brush tire model in (\ref{ref_model_tireFyFx_from_F}) and (\ref{ref_model_coupled_slip_fiala}) assuming lateral only slip and lumped tire forces. Next, we convert the estimated tire-fixed lateral force $F_{yf}$ to an estimated body-fixed lateral force $F_{1y}$ following the geometric relationship for the front axle in (\ref{controller_body_to_tire_forces}). The key step is to compute the rear lateral force $F_{2y}$ required to track $\Tilde{r}$ given the estimated front lateral force. We implement the following control law:
\begin{equation}
    F_{2y} = \frac{-\Tilde{M}_z+aF_{1y}+K_{rsat}e_r}{b},
    \label{controller_deltaf_sat_F2y}
\end{equation}
which uses proportional feedback $K_{rsat}$ to track the reference yaw rate. Substituting this equation into the expression for $\dot{e}_r$ in (\ref{controller_define_erdot_euydot}), there are stable error dynamics on yaw rate
\begin{equation}
    \dot{e}_r = \frac{K_{rsat}e_r}{I_z}
    \label{controller_deltaf_sat_er}
\end{equation}
if $K_{rsat} < 0$. For this work, we choose a value of $K_{rsat} = -12000$ Nm $\cdot$ s. The final step is to convert $F_{2y}$ to a rear steering command $\delta_r$ by following the calculations in (\ref{controller_body_to_tire_forces}) and (\ref{controller_steering_from_slip}).

\subsubsection{Acceleration at Driver's Seat Position}
For accurate vestibular feedback, the yaw rate and lateral acceleration should match the reference values at the position of the driver's seat. Following principles of rigid body dynamics, the rotation rate is the same everywhere on the vehicle, and the lateral acceleration at a location $\Delta x$ in front of and $\Delta y$ to the left of the center of mass is
\begin{equation}
    a_{ydriver} = a_y+\dot{r}\Delta x - r^2 \Delta y.
    \label{controller_ay_driver}
\end{equation}
We derive this kinematic relationship in the appendix. By tracking $\Tilde{r}$ and $\Tilde{a}_y$ at the center of mass, we also track lateral acceleration at the driver's position, since $a_{ydriver}$ only depends on $a_y$, $r$, and $\dot{r}$, which are tracked by our controller.

\subsection{Haptic Feedback}
\label{section_haptic_feedback}
In addition to visual and vestibular feedback, drivers use the torque felt on the steering wheel to understand their vehicle's response to steering inputs and the conditions of the road. In this section, we describe the design of a steering force feedback system that transmits realistic road feel to drivers corresponding to the high speed reference model. The haptic steering system in our ViL simulator follows a similar structure to the artificial steering feel model in SBW vehicles presented by Balachandran and Gerdes \cite{balachandran2014designing}. Here, the single track model shown in Fig. \ref{figure_single_track_model} provides the relevant dynamics for our force feedback model. 

To replicate the haptic feedback of a conventional road vehicle that mechanically links the steering wheel with the road wheels, the force feedback motor applies the following torque on the hand wheel:
\begin{equation}
    \tau_{hw} = \tau_{damp}+\tau_{inertia}+W_f(\Tilde{\alpha}_f)(\tau_{align}(\Tilde{\alpha}_f)+\tau_{jack}),
    \label{FFB_torque_overall}
\end{equation}
where $W_f$ and $\tau_{align}$ are both functions of the reference front slip angle $\Tilde{\alpha}_f$, which is calculated using the first expression in (\ref{controller_steering_from_slip}) with values in the reference model. 

The first two terms in (\ref{FFB_torque_overall}) replicate the inherent damping and inertia of a steering column and rack in a typical vehicle. The third term multiplies aligning and jacking torques caused by forces on the road wheels with a weighting function $W_f$, which emulates a power steering assist system. The weighting function $W_f$ is a Gaussian function centered with unity value at $\Tilde{\alpha}_f = 0$ and quickly dropping off to 0.2 with increasing slip angle magnitude, which mimics power steering. The aligning torque $\tau_{align}$ captures the effect of the front tires' lateral forces, tending to self-center the wheels at high speeds. Together with the jacking torque $\tau_{jack}$, which is a function of the driver's steering wheel angle, this last term gives the driver a feel for the dynamics of the reference vehicle through slip angles and tire forces at the front axle. Complete details of the artificial steering feel model can be found in \cite{balachandran2014designing}.

\section{Experiments}
\label{experiments}

We conducted experiments on a 110 m by 200 m skid pad at Thunderhill Raceway Park in Willows, CA. We tested our system with double lane change and highway weaving maneuvers that are difficult for drivers to navigate without multisensory feedback. Most of the plots below show data as a function of forward distance along the virtual road $s$. We have plotted relevant vehicle states, steering inputs, and steering wheel torque to demonstrate the multisensory feedback experienced by the driver in each experiment and the controller used to emulate high speed dynamics.


\subsection{Yaw Motion Thresholds}
In the analysis of our experimental data, models of the human vestibular system are useful for quantifying how well we reproduce rotations on the vehicle. Specifically, we investigate whether drivers can detect errors between the reference yaw rate and the yaw rate felt in the car. Nesti \textit{et al.} present a useful model for computing yaw rate perception thresholds based on human subjects studies in which participants were tasked with discriminating between yaw motions of different amplitudes using both visual and vestibular cues \cite{nesti2015human}. They demonstrate a nonlinear increase in yaw rate detection threshold with reference signal amplitude that is consistent with other studies of yaw perception, for example the findings of Mallery \textit{et al.} who tested participants with vestibular only cues \cite{mallery2010human}. Bruschetta \textit{et al.} use the model presented by Nesti \textit{et al.} to validate an MPC-based motion cueing approach on a moving platform driving simulator \cite{bruschetta2017fast}. Although Valente Pais \textit{et al.} find that humans may consider visual and vestibular yaw motion to be coherent up to a fairly wide 10 deg/s of divergence \cite{valente2010perception}, we follow the example in \cite{bruschetta2017fast} and use Nesti \textit{et al.}'s more conservative yaw rate threshold model.

\subsection{Double Lane Change Experiment}
\begin{figure*}[thpb]
    \centering
    \includegraphics[width=\linewidth, trim={3cm 1cm 2cm 2cm}, clip]{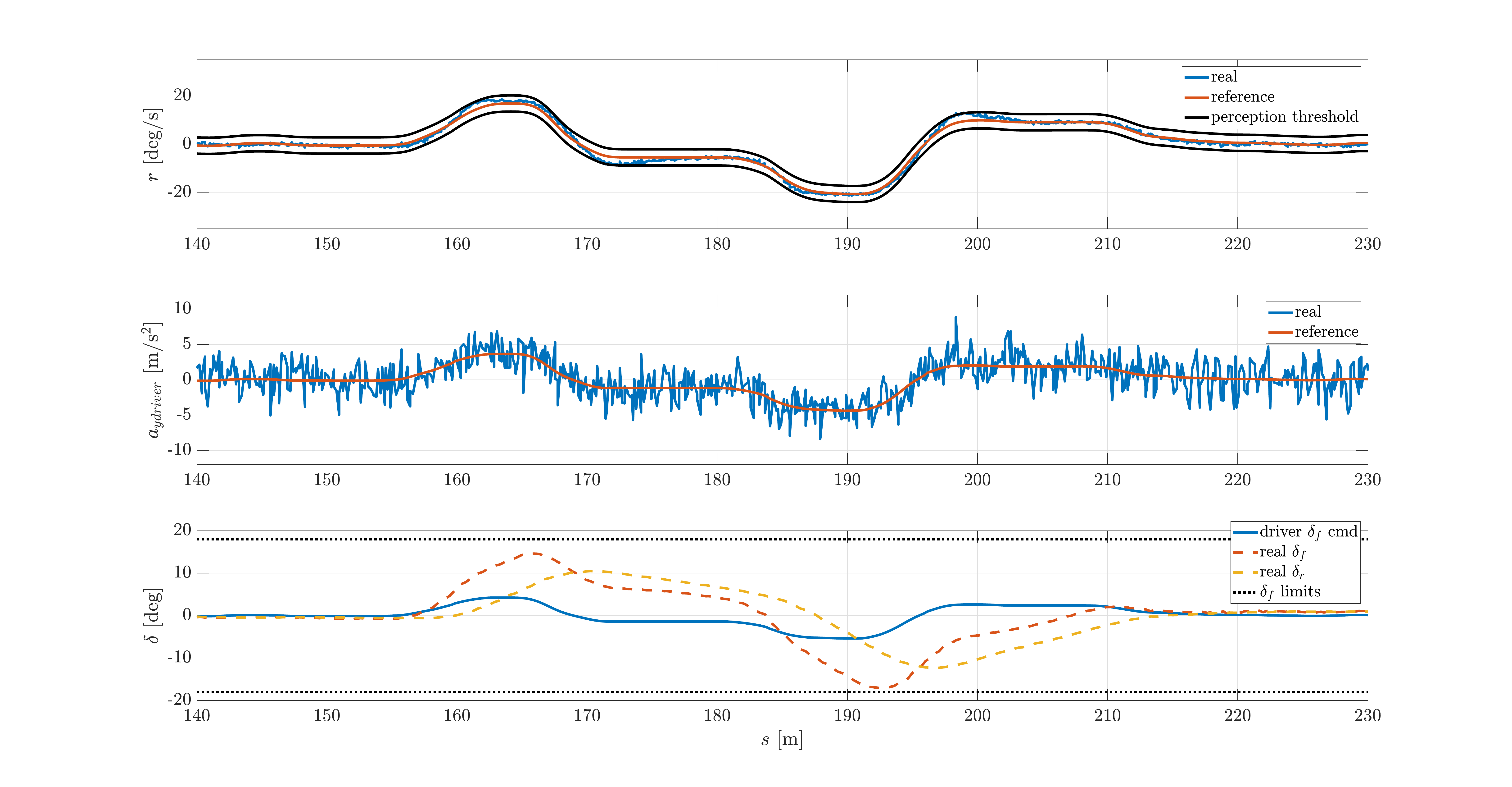}
    \caption{Yaw rate (top), lateral acceleration (middle), and steering (bottom) data from a double lane change at 30 mph with speed scaled by $f = 2$.}
    \label{figure_DLC20mph_timeseries}
\end{figure*}

In our first experiment, the driver completes an ISO 3888-2 double lane change, which requires two rapid lane changes over a 61 m course \cite{ISO3888}. We expect to see high amplitudes of yaw rate and lateral acceleration that change direction quickly during the maneuver and four-wheel steering on X1 that tracks these reference signals closely. Fig. \ref{figure_DLC20mph_timeseries}. shows data from a double lane change driven at a reference speed $\Tilde{u}_x =$ 30 mph with a speed scaling factor $f =$ 2 (and thus the real vehicle's speed $u_x =$ 15 mph). This experiment demonstrates the performance of our platform at high yaw rates and lateral accelerations, as driving the double lane change at 30 mph requires relatively aggressive steering inputs from the driver.

The top subplot shows the reference and measured yaw rates alongside the yaw detection threshold derived from Nesti \textit{et al.} of 3.35 deg/s based on the reference signal amplitude. Aside from a few brief time steps where yaw rate error slightly exceeds detection thresholds, the yaw rate felt by the driver stays within the bounds predicted by the perception model, emulating rotations up to 20.6 deg/s. The middle subplot shows lateral acceleration tracking at the driver's position on the car, computed with (\ref{controller_ay_driver}). The mean of the measured lateral acceleration signal tracks the reference signal well up to a maximum value of 4.4 m/s$^2$ and superimposes higher frequency content, which is analyzed in more detail below. The bottom subplot shows the driver's commanded steering angle overlaid with the front and rear steering angles sent to X1. Both axles tend to steer in the same direction and with greater magnitude than the driver's steering command throughout the maneuver, resulting in the exaggerated lateral motion needed to track the reference lateral acceleration at the slower speed on the real vehicle. The difference between the front and rear steering angles creates a positive yaw moment for leftward rotations and negative for rightward rotations, enabling the vehicle to track the reference yaw motion.

The front axle gets close to its maximum steering angle of $18^{\circ}$ at approximately $s =$ 193 m. The front steering geometry on X1 limits the amount of lateral acceleration that can be achieved during tests, as both axles need to increase steering angle to match the lateral motion required for acceleration tracking. We can widen this limitation by increasing the stiffness or adhesion of the tires (for example, high performance tires on race cars have friction coefficients upwards of $\mu \approx$ 1.3-1.6) or using a four-wheel steer vehicle with greater available steering angle on the front axle. In contrast to efforts needed to overcome the acceleration limits of moving platform simulators, which typically require the construction of larger facilities, these modifications to X1 are relatively cheap.

\subsection{Highway Weaving Experiment}
\begin{figure*}[thpb]
    \centering
    \includegraphics[width=\linewidth, trim={3cm 1cm 2cm 2cm}, clip]{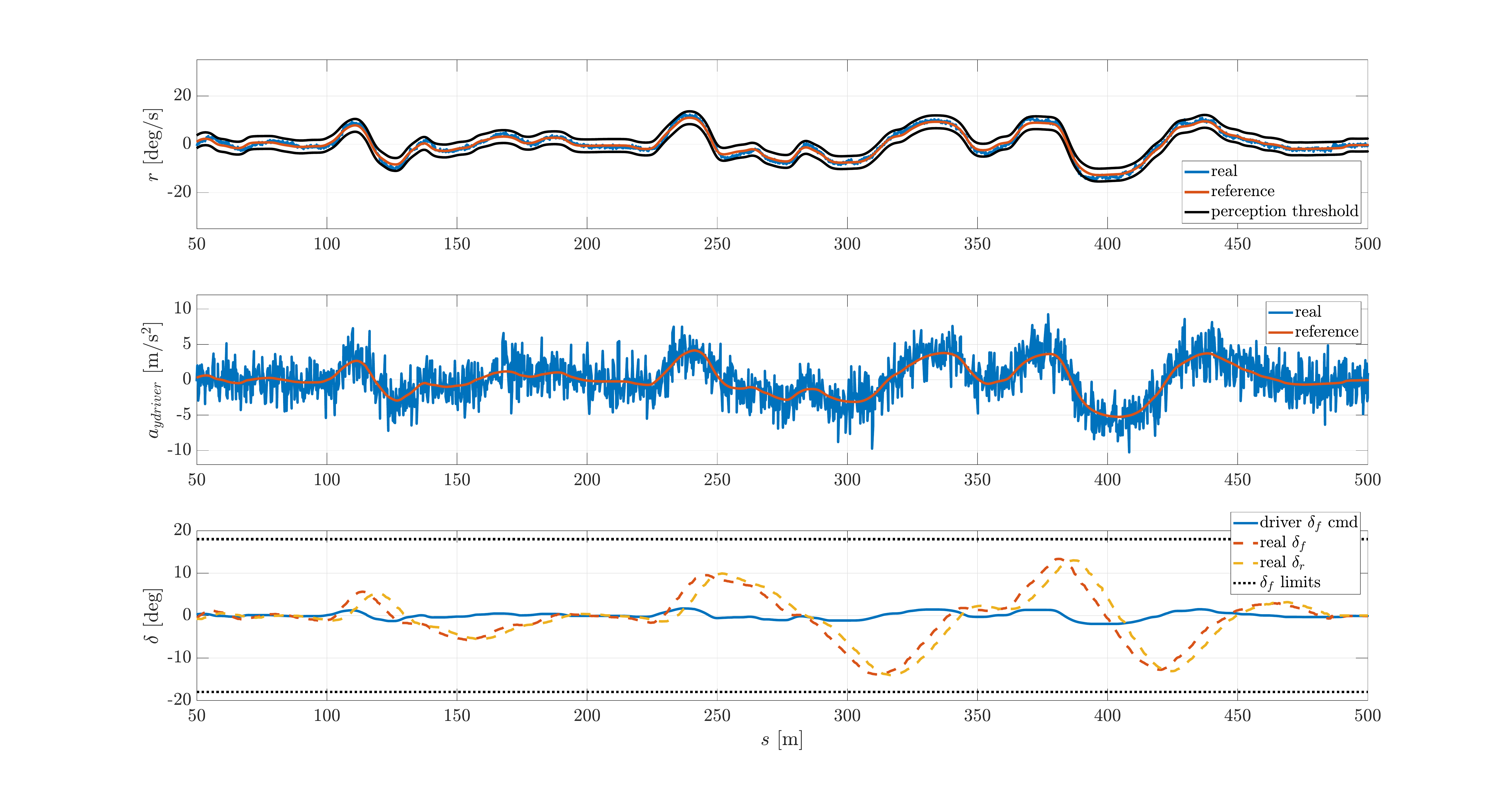}
    \caption{Yaw rate (top), lateral acceleration (middle), and steering (bottom) data from a highway weaving experiment at 60 mph with speed scaled by $f = 3$.}
    \label{figure_wiggle60mph_timeseries}
\end{figure*}

We additionally demonstrate the high speed emulation method in a much faster highway setting by running a test at a reference speed $\Tilde{u}_x =$ 60 mph with a speed scaling factor $f =$ 3. Here, the driver switches between two lanes as if they were aggressively weaving between traffic on the highway. We display data from this experiment in Fig. \ref{figure_wiggle60mph_timeseries}.

Similar to the double lane change experiment, the measured yaw rate stays within the perception threshold computed with Nesti \textit{et al.}'s model of 2.65 deg/s for the vast majority of the test. With higher speed traffic weaving, the vehicle tends to rotate less during lane changes, achieving a maximum yaw rate of 12.8 deg/s. The vehicle tracks the reference lateral acceleration closely, reproducing accelerations up to 5.25 m/s$^2$. The steering angle plot shows a similar sequence of control inputs on X1, where the front and rear axles steer in the same direction but with asymmetry in values that reproduces the reference yaw motion. Looking at the x-axis, we drive at a perceived speed of 60 mph over 450 m of longitudinal distance, expanding the effective testing area of the skid pad. 

\begin{figure}[thpb]
    \centering
    \includegraphics[width=\linewidth, trim={4.5cm 0.5cm 6cm 1cm}, clip]{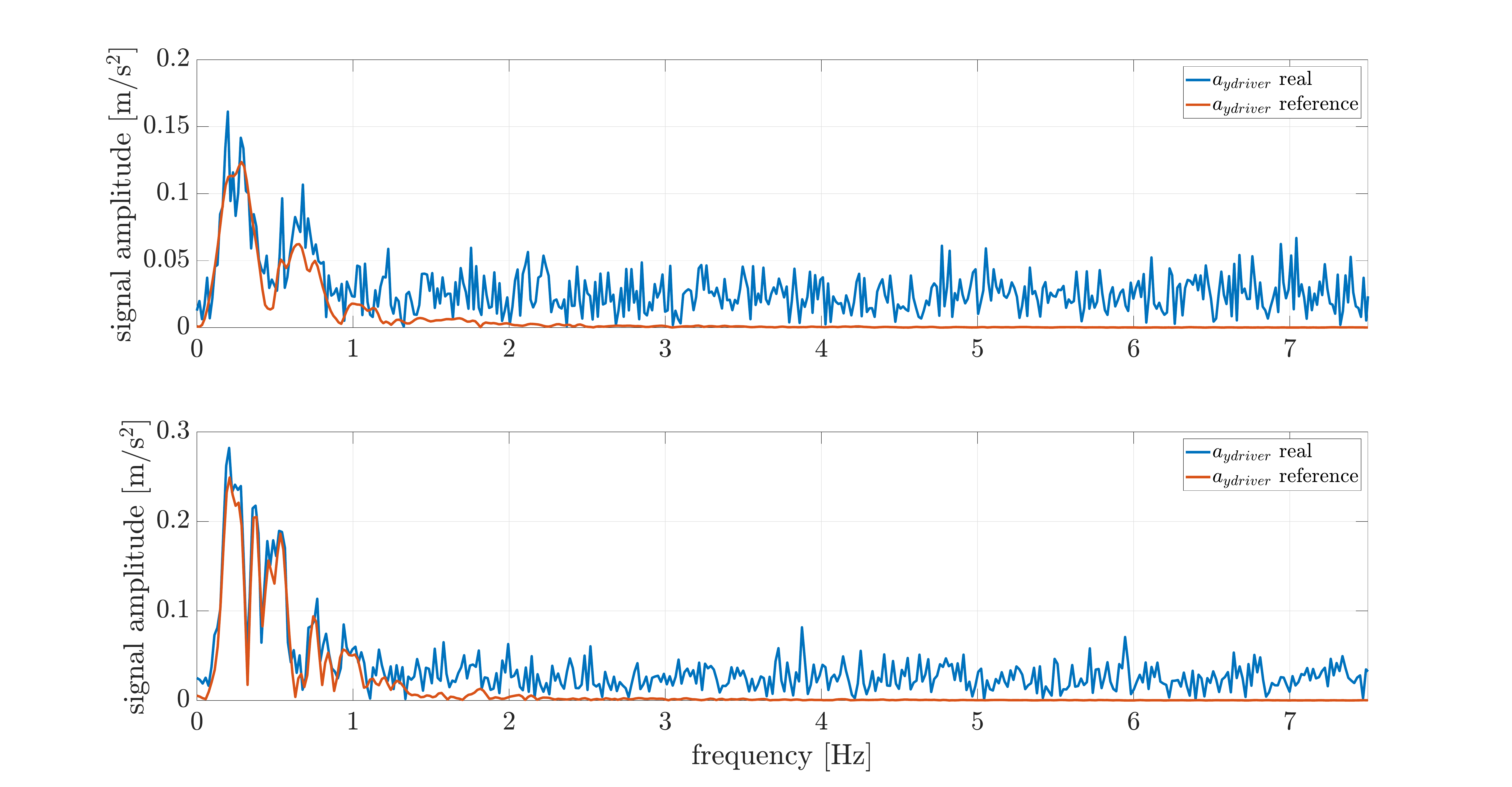}
    \caption{Frequency analysis of real and reference lateral acceleration signals for the 30 mph double lane change (top) and 60 mph highway weaving experiment (bottom).}
    \label{figure_combined_ay_fft}
\end{figure}

\subsection{Frequency Analysis of Lateral Acceleration Data}
We also analyze lateral acceleration data in the frequency domain through a discrete Fourier transform. Fig. \ref{figure_combined_ay_fft} plots the amplitudes of each transformed signal versus frequency for the double lane change and the highway traffic weaving experiments.

In this representation, the low frequency behavior of the reference and measured data closely match. In the double lane change on the top subplot, there are larger acceleration amplitudes around 0.3 Hz and 0.7 Hz. For the highway weaving experiment in the bottom subplot with slightly higher frequency lane changes by the driver, frequency content matches well between the two signals up to 1 Hz. Hence, we reproduce the lower frequency lateral acceleration associated with the maneuvers themselves on the vehicle. Above approximately 1 Hz, although the reference model does not compute higher frequency content, the vehicle measures nonzero values in lateral acceleration. This frequency content represents high frequency accelerations associated with the rough road surface and the resulting suspension motion that we don't include in the reference model. The driver experiences the low frequency accelerations associated with the reference model superimposed with the high frequency accelerations one expects while driving a real vehicle. This high frequency content is important for immersing drivers in a simulator, especially when emulating highway speed motion.

\begin{figure}[thpb]
    \centering
    \includegraphics[width=\linewidth, trim={5.2cm 0.5cm 5.3cm 1cm}, clip]{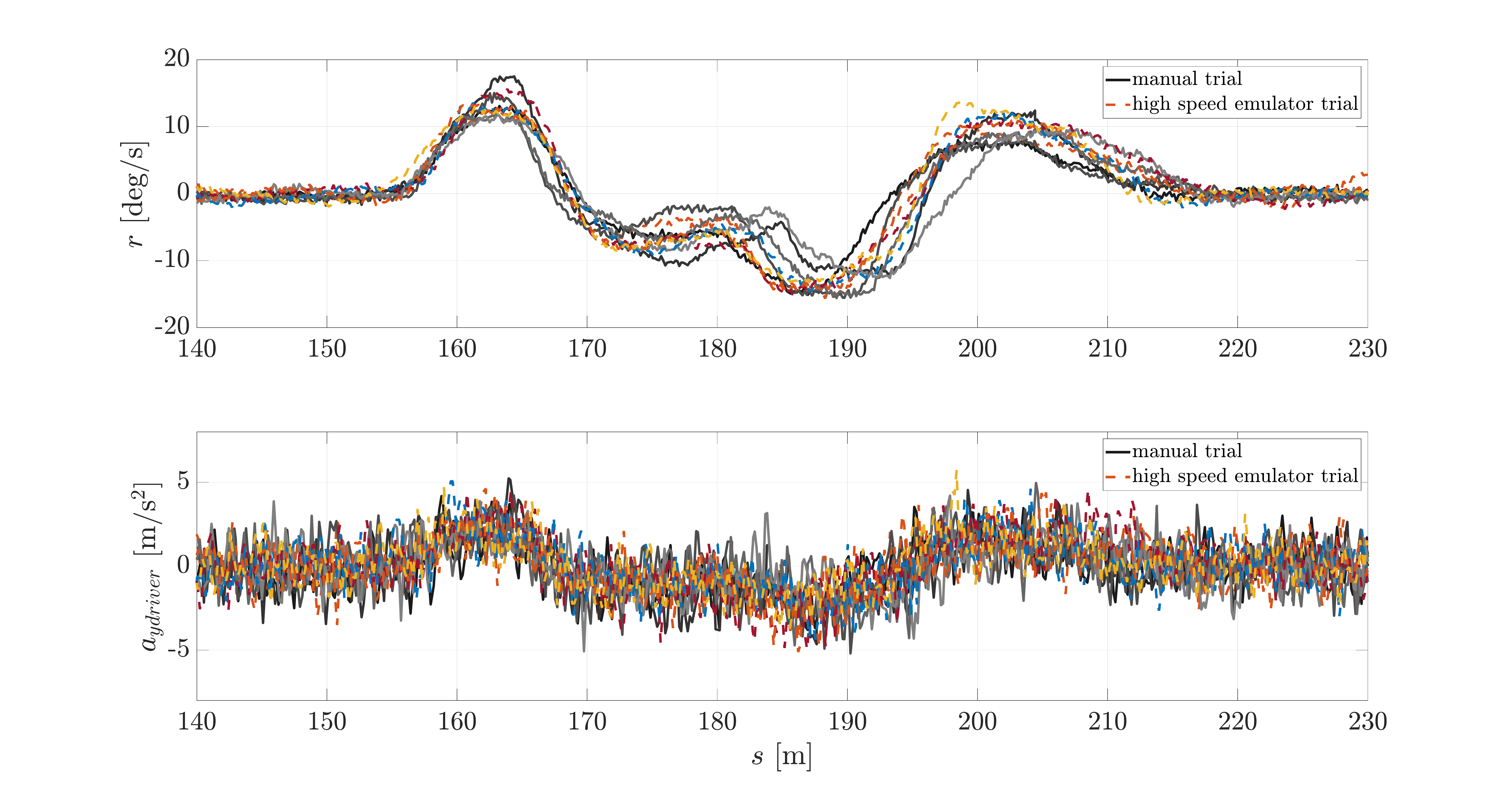}
    \caption{Yaw rate (top) and lateral acceleration (bottom) overlaid for five manual and five high speed emulation double lane changes at 20-22 mph with speed scaled by $f = 2$. Manual trials are grey solid lines, and high speed emulation trials are colored dashed lines.}
    \label{figure_5compare_r_ay}
\end{figure}

\begin{figure}[thpb]
    \centering
    \includegraphics[width=\linewidth, trim={5.5cm 0.5cm 5.3cm 1cm}, clip]{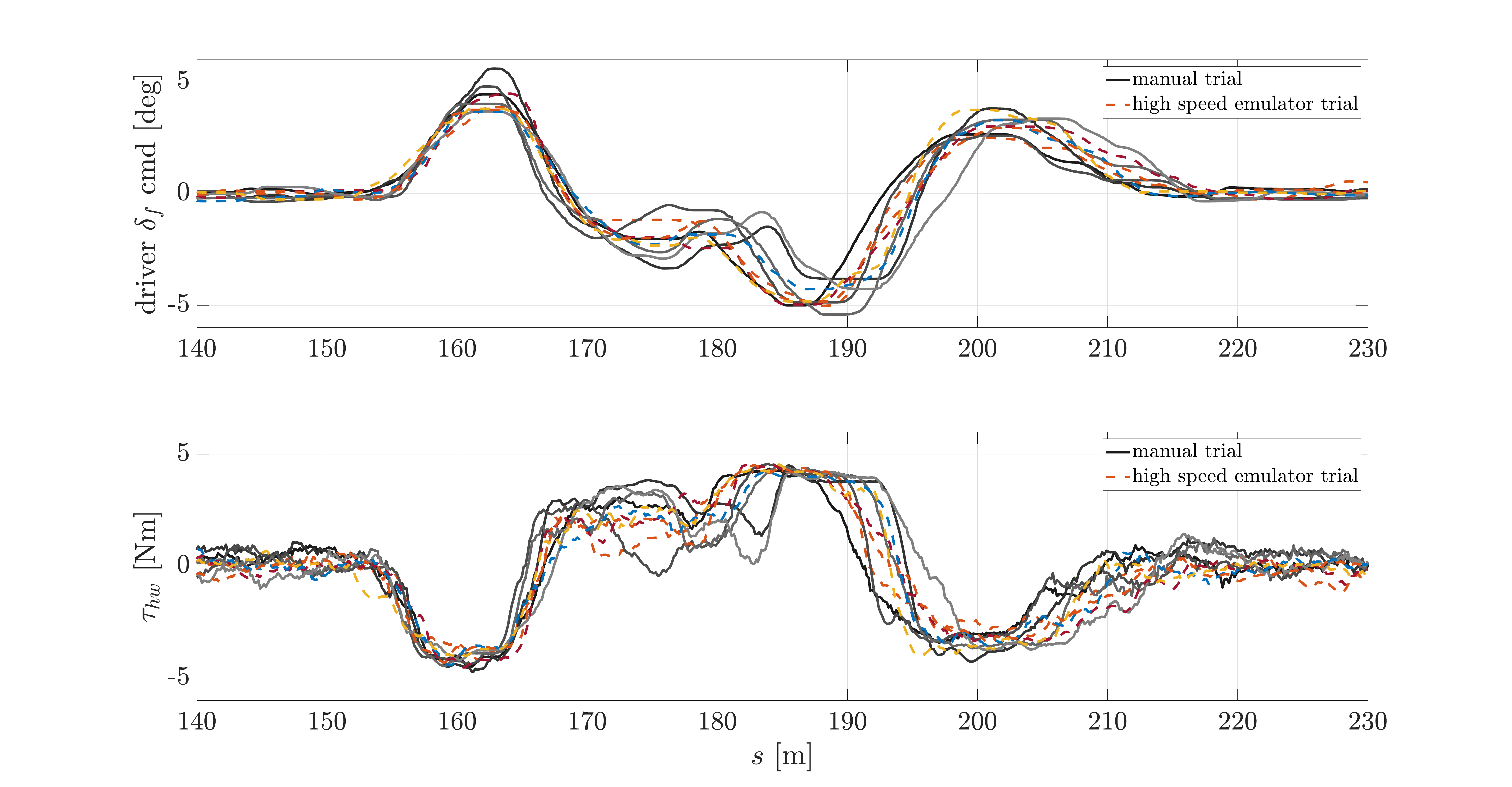}
    \caption{Driver steering input (top) and steering wheel haptic feedback (bottom) overlaid for the same five manual and five high speed emulation double lane changes.}
    \label{figure_5compare_delta_FFB}
\end{figure}

\subsection{Comparison With Manual Tests}
Our final set of tests compares the high speed emulation method with an equivalent manual driving experience. We use the ISO double lane change as a benchmark for these experiments. We expect similar driving behavior and sensory feedback between the two conditions, indicating that they are indistinguishable from one another to the driver.

A driver first completed five trials of the double lane change with $\Tilde{u}_x$ between 20-22 mph and $f = 2$ in the high speed emulation condition. Using a lower speed for these tests reduces the demanded lateral acceleration, ensuring that the front steering angle isn't saturated. The same driver then completed five trials of the double lane change at 20-22 mph controlling the ViL simulator fully manually. In the manual driving condition, the virtual vehicle's motion corresponds exactly to the real vehicle's motion, the driver's steering commands directly control X1 using front-wheel steering only, and the haptic feedback model on the steering wheel uses the front slip angle $\alpha_f$ measured on the vehicle. Figs. \ref{figure_5compare_r_ay} and \ref{figure_5compare_delta_FFB} show measured data on X1 from this set of experiments.

Across the plots, there is overlap between the yaw rate, lateral acceleration, steering command, and steering wheel torque measured in the high speed emulation maneuvers and the manual maneuvers. The range in yaw rate and lateral acceleration experienced by the driver through the maneuver doesn't vary significantly between the two conditions. In general, the spread in data between the trials tends to be greater for the manual condition. This emerges most clearly in the steering command plot at the top of Fig. \ref{figure_5compare_delta_FFB}, where there is higher variance in the driver's control input from $s = 170$ m to 190 m than in the high speed emulation trials. The variance in vestibular and haptic feedback data for the manual trials corresponds to this spread in control input data, indicating that the dynamics of the vehicle and the force feedback model used to emulate high speeds closely match those of a manually driven vehicle at the same speed. The haptic feedback model in particular provides very similar steering wheel torques at the points of greatest force feedback at 160 m and 185 m along the path. The small but visible difference in driving behavior could be caused by a number of factors, including the trial ordering, and would require a study with more participants to discern underlying causes. The lack of clear clustering of data between the two conditions, however, demonstrates the realism of our high speed emulation approach through its similarity to manual experiments driven at the same speed.

\subsection{Discussion}
In this paper, we  demonstrate a method for generating haptic and vestibular cues that meet objective performance measures. While it is common to conduct human subjects studies to validate novel driving simulator designs, objective criteria set clear standards for evaluating drivers' level of immersion that can be applied even if the simulator design changes. Underscoring this approach, recent work by Grottoli \textit{et al.} \cite{grottoli2019objective} and by Biemelt \textit{et al.} \cite{biemelt2019objective} use objective metrics of vestibular feedback quality and known perception thresholds to assess the effectiveness of different motion cueing algorithms. In our method, the vestibular feedback experienced by the driver is verified objectively by comparing the difference between the real and reference yaw rate to known yaw perception thresholds derived by Nesti \textit{et al.} from human subjects studies \cite{nesti2015human}. Motion perception thresholds play an important role in the development and evaluation of many motion cueing algorithms and are central to analysis of the vestibular feedback experienced by the driver in our high speed emulation method. Further, the haptic feedback is designed and tuned to match objective steering characteristics. Following the approach presented by Balachandran and Gerdes, the artificial steering feel model meets objective measures such as returnability, on-center feel, and steering sensitivity \cite{balachandran2014designing}. Ultimately, our system enables a wide range of immersive tests that may be too costly or unsafe to perform at full speeds. We show that objective measures for vestibular and haptic feedback are met in double lane change and highway weaving maneuvers at various speeds. Further testing, which can be objective or subjective, may be required to validate this simulation platform for a particular experiment.

\section{Conclusion}
\label{conclusion}
We have developed a novel ViL driving simulator using a four-wheel SBW vehicle that renders multisensory feedback to drivers in high speed driving conditions. Building on prior work in vehicle dynamics emulation and artificial steering feel, we demonstrate a method for simulating the dynamics of a faster moving vehicle and rendering the corresponding visual, vestibular, and haptic feedback to the driver. We tested this system with double lane change and highway traffic weaving experiments. These tests show that we can render yaw rates within drivers' perceptual thresholds, recreate accurate lateral accelerations across a broad frequency spectrum, and generate an equivalent driving experience while navigating the same double lane change at full speed. 

We note that there are a few limitations to this approach. By scaling the speed of the vehicle, we also scale the longitudinal acceleration, causing the driver to experience a mismatch between the longitudinal acceleration seen in the VR HMD and felt in their vestibular system. We run tests at relatively constant speeds, leaving this mismatch to only occur when speeding up to or slowing down from the target speed of the experiment. Additionally, the high speed emulation system does not easily handle steady-state lateral accelerations, as we cannot track a constant $\dot{u}_{ydes}$ on the vehicle for long before the wheels run out of additional steering angle needed to increase lateral velocity. We therefore run tests along straight roads involving transient lateral maneuvers at constant speeds. Within this set of maneuvers, however, we can simulate many relevant scenarios such as emergency double lane changes and vehicle overtaking on the highway. 

Our ViL platform is useful for safely studying driver behavior in extreme driving scenarios at a large range of speeds. In the future, we plan to study emergency highway lane changes in more detail to develop and test advanced driver assistance systems that can keep drivers safe despite the challenges of controlling a high speed vehicle and uncertain behavior from other road users.

\section*{Acknowledgments}
We would like to thank Alessandra Napoli, John Talbot, and Larry Cathey for help with experiments, Markus Steimle for insights on developing our ViL platform, and Virtual Test Drive for providing academic licenses and software support. Toyota Research Institute provided funds to support this work.

{\appendix[Derivation of Acceleration at Driver's Position]
From analysis of a rotating and translating rigid body, we obtain the following expression for the acceleration of a point $p$ on the body in a reference frame fixed to the body
\begin{equation}
    \boldsymbol{a_p} = \boldsymbol{a_{CoM}} + \boldsymbol{\dot{\omega}} \times \boldsymbol{r} + \boldsymbol{\omega} \times (\boldsymbol{\omega} \times \boldsymbol{r}),  
    \label{appendix_rigid_acceleration}
\end{equation}
where $\boldsymbol{a_{CoM}}$ is the acceleration vector at the center of mass, $\boldsymbol{\omega}$ is the rotational velocity vector, and $\boldsymbol{r}$ is the position of the point $p$ relative to the center of mass, all in body-fixed coordinates. We define these vectors for the planar vehicle model as
\begin{equation}
    \boldsymbol{a_{CoM}} = \begin{bmatrix} a_x \\ a_y \\ 0 \end{bmatrix}, ~~
    \boldsymbol{\omega} = \begin{bmatrix} 0 \\ 0 \\ r \end{bmatrix}, ~~
    \boldsymbol{r} = \begin{bmatrix} \Delta x \\ \Delta y \\ 0 \end{bmatrix}.
    \label{appendix_rigid_body_vector_defs}
\end{equation}
Using these definitions, the acceleration vector at the position $p$ in the frame of the vehicle is
\begin{equation}
    \boldsymbol{a_p} = \begin{bmatrix} a_x-\dot{r}\Delta y - r^2 \Delta x \\ a_y+\dot{r}\Delta x - r^2 \Delta y \\ 0 \end{bmatrix}.
    \label{appendix_aydriver}
\end{equation}}

\bibliographystyle{IEEEtran}
\bibliography{IEEEabrv, myBib}



\vfill

\end{document}